\documentclass[11pt,reqno]{amsart}

\usepackage{amssymb,amsmath,amsthm,amscd,latexsym,amsfonts}
\usepackage{mathtools}
\usepackage[T1]{fontenc}
\usepackage{graphicx,epstopdf}
\usepackage{graphicx}
\usepackage{xcolor}
\usepackage{cite}

\usepackage[a4paper,top=3cm, bottom=3cm, left=3.1cm, right=3.1cm]{geometry}

\newtheorem{thm}{Theorem}
\newtheorem{defn}{Definition}
\newtheorem{lemma}{Lemma}
\newtheorem{pro}{Proposition}
\newtheorem{rk}{Remark}

\newtheorem{ex}{Example}

\numberwithin{equation}{section} 

\newcommand{\s}{{\sigma}}

\newcommand{\bea}{\begin{eqnarray}}
\newcommand{\eea}{\end{eqnarray}}

\newcommand{\Z}{\mathbb{Z}}

\def\s{\sigma}

\def\s{\sigma}

\def\O{\Omega}

\def\s{\sigma}

\def\a{\alpha}

\def\O{\Omega}

\def\b{\beta}

\def\L{\Lambda}

\def\Z{\mathbb{Z}}

\def \L {\Lambda}

\def\c{\gamma}

\begin{document}
\title [Gradient Gibbs measures for SOS model]
{Gradient Gibbs measures of a SOS model on Cayley trees: 4-periodic boundary laws}

\author { F. H. Haydarov, U.A. Rozikov}

\
\address{F.\ H.\ Haydarov$^{a,b}$\begin{itemize}
\item[$^a$] National University of Uzbekistan,
Tashkent, Uzbekistan.
\item[$^b$] AKFA University, 1st Deadlock 10, Kukcha Darvoza, 100095, Tashkent, Uzbekistan.
\end{itemize}}
\email{haydarov\_imc@mail.ru}

\address{ U.Rozikov$^{c,b,d}$\begin{itemize}
 \item[$^c$] V.I.Romanovskiy Institute of Mathematics of Uzbek Academy of Sciences;
\item[$^b$] AKFA University, 1st Deadlock 10, Kukcha Darvoza, 100095, Tashkent, Uzbekistan;
\item[$^d$] Faculty of Mathematics, National University of Uzbekistan.
\end{itemize}}
\email{rozikovu@yandex.ru}

\begin{abstract}
 For SOS (solid-on-solid) model with external field and with spin values from the set of all integers,
 on a Cayley tree we give gradient Gibbs measures (GGMs).
Such a measure corresponds to a boundary law (a function defined on vertices of Cayley tree)
satisfying an infinite system of functional equations.
We give several concrete GGMs which correspond to periodic boundary laws.

\end{abstract}
\maketitle



\section{Introduction and known results}

The study of random field $\xi_x$ from a lattice graph $\mathbb L$ (usually $\mathbb Z^d$
or a Cayley tree $\Gamma^k$) to a measure space $(E, \mathcal E)$ is a
central component of ergodic theory and statistical physics.

In many classical models
from physics (e.g., the Ising model, the Potts model), $E$ is a finite set (i.e., with a finite underlying measure $\lambda$), and $\xi_x$
has a physical interpretation as the spin of a particle
at location $x$ in a crystal lattice.

Following \cite{BiKo}, \cite{BEvE}, \cite{FS97}, \cite{Ge}, \cite{HKLR}, \cite{HKa}, \cite{HKb}, \cite{KS}, \cite{Sh}, \cite{Z1}, let us
give basic definitions and some known facts related to (gradient) Gibbs measures.

{\bf $\sigma$-algebra, Hamiltonian.}
In general, $(E, \mathcal E)$ is a space with an infinite underlying measure $\lambda$ (i.e. $\mathbb L$ with counting measure),
where $\mathcal E$ is the Borel $\sigma$-algebra of $E$ and $\xi_x$ usually
has a physical interpretation as the spatial position of a particle at location $x$
in a lattice. In \cite{FS97} first such models were considered.

The prime examples of unbounded spin systems are harmonic
oscillators. Another example is the Ginzburg-Landau
interface model; which is obtained from the harmonic oscillators \cite{Ge}, \cite{Sh}.

Denote by $\Omega$ the set of functions from $\mathbb L$ to $E$, such a function also is called  a configuration.

Assume random field $(\xi_x)_{x\in \mathbb L}$ on $\Omega$ given as the projection onto the coordinate $x\in \mathbb L$:
$$\xi_x(\omega)=\omega(x)=\omega_x, \ \ \omega \in \Omega.$$

For $\Lambda\subset \mathbb L$, denote by $\mathcal F_\Lambda$
the smallest $\sigma$-algebra with respect to which $\xi_x$ is measurable for all $x\in \Lambda$. Write $\mathcal T_\Lambda=\mathcal F_{\mathbb L\setminus\Lambda}$.

A subset of $\Omega$,
is called a cylinder set if it belongs to $\mathcal F_\Lambda$ for some finite set $\Lambda \subset \mathbb L$.

Let $\mathcal F$ be the smallest
$\sigma$-algebra on $\Omega$ containing the cylinder sets.

Write $\mathcal T$ for the tail-$\sigma$-algebra, i.e., intersection of $\mathcal T_\Lambda$
over all finite subsets $\Lambda$ of $\mathbb L$ the sets in $\mathcal T$ are called tail-measurable sets.

Assume that we are given a family of measurable potential
functions $\Phi_\Lambda:\Omega\to \mathbb R\cup \{\infty\}$ (one for each finite subset $\Lambda$ of $\mathbb L$)
 each  $\Phi_\Lambda$ is $\mathcal F_\Lambda$
measurable.

For each finite subset $\Lambda$ of $\mathbb L$ define a Hamiltonian:
$$ H_\Lambda(\sigma)=\sum_{S\subset \mathbb L:\atop S\cap \Lambda\ne \emptyset}\Phi_{S}(\sigma),$$
 where the sum is taken over finite subsets $S$.

{\bf Gibbs Measures.}
To define Gibbs measures and gradient Gibbs measures, we will need some additional
notation \cite{Ge}, \cite{Sh}.

Let $(X, \mathcal X)$ and $(Y, \mathcal Y)$ be general measure spaces.

A function $\pi : \mathcal X\times Y\to [0, \infty]$ is called a probability kernel from $(Y, \mathcal Y)$ to $(X, \mathcal X)$ if
\begin{itemize}
\item[1.] $\pi(\cdot | y)$ is a probability measure on $(X, \mathcal X)$ for each fixed $y\in Y$, and
\item[2.] $\pi(A | \cdot)$ is $\mathcal Y$-measurable for each fixed $A\in \mathcal X$.
\end{itemize}

Such a kernel maps each measure $\mu$, on $(Y, \mathcal Y)$ to a measure $\mu\pi$ on $(X, \mathcal X)$ by
$$\mu\pi(A)=\int \pi(A | \cdot)d\mu$$

The following is a probability kernel from $(\Omega, \mathcal T_\Lambda)$ to $(\Omega, \mathcal F)$:
$$\gamma_\Lambda(A,\omega)=Z_\Lambda(\omega)^{-1}\int\exp(-H_\Lambda(\sigma_\Lambda\omega_{\Lambda^c}))
\mathbf{1}_A(\sigma_\Lambda\omega_{\Lambda^c})\nu^{\otimes\Lambda}(d\sigma_\Lambda),$$
where $\nu=\{\nu(i)>0, i\in E\}$ is a counting measure.

A configuration $\sigma$ has finite energy if $ \Phi_\Lambda(\sigma) < \infty$ for all finite $\Lambda$. Moreover, $\sigma$ is $\Phi$-admissible
if each $Z_\Lambda(\sigma)$ is finite and non-zero.

Given a measure $\mu$ on $(\Omega, \mathcal F)$,  define a new
measure $\mu\gamma_\Lambda$ by
$$\mu\gamma_\Lambda(A)=\int\gamma_\Lambda(A, \cdot)d\mu$$

\begin{defn} A probability measure $\mu$ on $(\Omega, \mathcal F)$ is called a Gibbs measure if $\mu$ is supported on the set of $\Phi$-admissible configurations in $\Omega$ and for all finite subset $\Lambda$ we have
$$\mu \gamma_\Lambda=\mu.$$
\end{defn}
{\bf Gradient Gibbs measure.} For any configuration $\omega = (\omega(x))_{x \in \mathbb L} \in E^\mathbb L$ and edge $b = \langle x,y \rangle$ of $\mathbb L$
the \textit{difference} along the edge $b$ is given by $\nabla \omega_b = \omega_y - \omega_x$ and $\nabla \omega$ is called the \textit{gradient field} of $\omega$.

The gradient spin variables are now defined by $\eta_{\langle x,y \rangle} = \omega_y - \omega_x$ for each $\langle x,y \rangle$.

The space of \textit{gradient configurations} denoted by $\O^\nabla$. The measurable structure on the space $\Omega^{\nabla}$ is given by $\sigma$-algebra $$\mathcal{F}^\nabla:=\sigma(\{ \eta_b \, \vert \, b \in \mathbb L \}).$$
Note that $\mathcal F^\nabla$ is the subset of
$\mathcal F$ containing those sets that are invariant under translations $\omega\to \omega +c$ for $c\in E$.

Similarly, we define
$$\mathcal T^\nabla_\Lambda=\mathcal T_\Lambda \cap \mathcal{F}^\nabla, \ \ \mathcal F^\nabla_\Lambda=\mathcal F_\Lambda \cap \mathcal{F}^\nabla.$$

Let $\Phi$ be a translation invariant
gradient potential. Since, given any $A\in \mathcal F^\nabla$, the kernels $\gamma^\Phi_\Lambda(A, \omega)$ are $\mathcal F^\nabla$-measurable
functions of $\omega$, it follows that the kernel sends a given measure $\mu$ on $(\Omega, \mathcal F^\nabla)$ to another
measure $\mu\gamma_\Lambda^\Phi$ on $(\Omega, \mathcal F^\nabla)$.

\begin{defn} A measure $\mu$ on $(\Omega, \mathcal F^\nabla)$ is called a gradient Gibbs measure
if for all finite subset $\Lambda$ we have
$$\mu \gamma^\Phi_\Lambda=\mu.$$
\end{defn}
Note that, if $\mu$ is a Gibbs measure on $(\Omega, \mathcal F)$, then its restriction to $\mathcal F^\nabla$ is a gradient
Gibbs measure.

A gradient Gibbs measure is said to be localized or smooth if it arises
as the restriction of a Gibbs measure in this way. Otherwise, it is non-localized or
rough.

It is known\cite{Ge}, \cite[Theorem 8.19.]{FV} that
many natural Gibbs measures on $\mathbb Z^d$ are rough when $d\in \{1, 2\}$.

{\bf Construction of gradient Gibbs measure on Cayley trees}. Following \cite{KS}
we consider models where spin-configuration $\omega$ is a function from the
vertices of the Cayley tree $\Gamma^k=(V, \vec L)$  to the set $E= \Z$, where
$V$ is the set of vertices and $\vec L$ is the set of oriented edges (bonds) of the tree
(see Chapter 1 of \cite{Ro} for properties of the Cayley tree).

For nearest-neighboring (n.n.) interaction  potential $\Phi=(\Phi_b)_b$, where
$b=\langle x,y \rangle$ is an edge,  define symmetric transfer matrices $Q_b$ by
\begin{equation}\label{Qd}
Q_b(\omega_b) = e^{- \big(\Phi_b(\omega_b) + | \partial x|^{-1} \Phi_{\{x\}}(\omega_x) + |\partial y |^{-1} \Phi_{\{y\}} (\omega_y) \big)},
\end{equation}
where $\partial x$ is the set of all nearest-neighbors of $x$ and $|S|$ denotes the number of elements of the set $S$.

Define the Markov (Gibbsian) specification as
$$
\gamma_\Lambda^\Phi(\sigma_\Lambda = \omega_\Lambda | \omega) = (Z_\Lambda^\Phi)(\omega)^{-1} \prod_{b \cap \Lambda \neq \emptyset} Q_b(\omega_b).
$$

If for any bond $b=\langle x,y \rangle$ the transfer operator $Q_b(\omega_b)$ is
a function of gradient spin variable $\zeta_b=\omega_y-\omega_x$ then the underlying potential $\Phi$ is called
a \textit{gradient interaction potential}.

 \emph{Boundary laws} (see \cite{Z1}) which allow to describe the set $\mathcal{G}(\gamma)$ of all Gibbs measures (that are Markov chains on trees).

\begin{defn}
	A family of vectors $\{ l_{xy} \}_{\langle x,y \rangle \in \vec L}$ with $l_{xy}=\left(l_{xy}(i): i\in \Z\right) \in (0, \infty)^\Z$ is called a {\em boundary law for the transfer operators $\{ Q_b\}_{b \in \vec L}$} if for each $\langle x,y \rangle \in \vec L$ there exists a constant  $c_{xy}>0$ such that the consistency equation
	\begin{equation}\label{eq:bl}
	l_{xy}(i) = c_{xy} \prod_{z \in \partial x \setminus \{y \}} \sum_{j \in \Z} Q_{zx}(i,j) l_{zx}(j)
	\end{equation}
	holds for every $i \in \Z$.

A boundary law is called {\em $q$-periodic} if $l_{xy} (i + q) = l_{xy}(i)$ for every oriented edge $\langle x,y \rangle \in \vec L$ and each $i \in \Z$.
	
\end{defn}

It is known that there is a one-to-one correspondence between boundary laws
and tree-indexed Markov chains if the boundary laws are {\em normalisable} in the sense of Zachary \cite{Z1}:

\begin{defn} A boundary law $l$ is said to be {\em normalisable} if and only if
\begin{equation}\label{Norm}
\sum_{i \in \Z} \Big( \prod_{z \in \partial x} \sum_{j \in \Z} Q_{zx}(i,j) l_{zx}(j) \Big) < \infty
\end{equation} at any $x \in V$.
\end{defn}
For any $\Lambda \subset V$ we define its outer boundary as
\begin{equation*}
\partial \Lambda := \{ x \notin \Lambda : \langle x,y\rangle \, \mbox{ for some } \, y \in \Lambda\}.
\end{equation*}
The correspondence now reads the following:

\begin{thm} \cite{Z1}
For any Markov specification $\gamma$ with associated family of transfer matrices $(Q_b)_{b \in L}$  we have
\begin{enumerate}
\item Each {\it normalisable} boundary law $(l_{xy})_{x,y}$ for $(Q_b)_{b \in L}$ defines a unique tree-indexed Markov chain $\mu \in \mathcal{G}(\gamma)$ via the equation given for any connected set $\Lambda \subset V$
\begin{equation}\label{BoundMC}
\mu(\sigma_{\Lambda \cup \partial \Lambda}=\omega_{\Lambda \cup \partial \Lambda}) = (Z_\Lambda)^{-1} \prod_{y \in \partial \Lambda} l_{y y_\Lambda}(\omega_y) \prod_{b \cap \Lambda \neq \emptyset} Q_b(\omega_b),
\end{equation}
where for any $y \in \partial \Lambda$, $y_\Lambda$ denotes the unique $n.n.$ of $y$ in $\Lambda$.
\item Conversely, every tree-indexed Markov chain $\mu \in \mathcal{G}(\gamma)$ admits a representation of the form (\ref{BoundMC}) in terms of a {\it normalisable} boundary law (unique up to a constant positive factor).
\end{enumerate}
\end{thm}

The Markov chain $\mu$ defined in \eqref{BoundMC} has the transition probabilities
\begin{equation}\label{last}
P_{xy}(i,j)=\mu(\s_y = j \mid \s_x =i)
= \frac{l_{yx}(j) Q_{yx}(j, i)}{\sum_s l_{yx}(s) Q_{yx}(s, i)}.
\end{equation}
The expressions \eqref{last}  may exist even in situations where the underlying boundary
law $(l_{xy})_{x,y}$ is not normalisable. However, the Markov chain given by (\ref{last}), in general, does not have an invariant probability measure.
Therefore in \cite{HKLR},  \cite{HKa}, \cite{HKb}, \cite{KS}
some non-normalisable boundary laws are used to give gradient Gibbs measures.

 Now we give some results of above mentioned papers. Consider a model on Cayley tree $\Gamma^k=(V, \vec L)$, where the spin takes values in
the set of all integer numbers $\mathbb Z$. The set of all configurations is $\Omega:=\mathbb Z^V$.

For $\Lambda\subset V$, fix  a site $w \in \Lambda$. If the boundary law $l$ is assumed to be $q$-periodic, then take $s \in \mathbb{Z}_q=\{0,1,\dots,q-1\}$ and define probability measure $\nu_{w,s}$ on $\mathbb{Z}^{\{b \in \vec L \mid b \subset \Lambda\}}$ by
$$
\nu_{w,s}(\eta_{\Lambda \cup \partial \Lambda}=\zeta_{\Lambda \cup \partial \Lambda})=
$$
$$Z^\Lambda_{w,s}\prod_{y \in \partial \Lambda} l_{yy_\L}\Bigl (T_q(
s+\sum_{b\in \Gamma(w,y)}\zeta_b)
\Bigr) \prod_{b \cap \Lambda \neq \emptyset}Q_b(\zeta_b),
$$
where $Z^\Lambda_{w,s}$ is a normalization constant, $\Gamma(w,y)$ is the unique path from $w$ to $y$
and $T_q: \mathbb{Z} \rightarrow \mathbb{Z}_q$ denotes the coset projection.

\begin{thm} \cite{KS}
	Let $(l_{<xy>})_{<x,y> \in \vec L}$ be any $q$
-periodic boundary law to some gradient interaction potential.
Fix any site $w \in V$ and any class label $s \in \mathbb{Z}_q$. Then
$$	\nu_{w,s}(\eta_{\Lambda \cup \partial \Lambda}=\zeta_{\L\cup\partial\L})
	=$$
\begin{equation}
Z^\Lambda_{w,s} \prod_{y \in \partial \Lambda} l_{yy_\L}\Bigl (T_q(
	s+\sum_{b\in \Gamma(w,y)}\zeta_b)
	\Bigr) \prod_{b \cap \Lambda \neq \emptyset}
	Q_b(\zeta_b),
	\end{equation}
gives a consistent family of probability measures on the gradient space $\Omega^\nabla$.
Here $\Lambda$ with $w \in  \L \subset V$ is any finite connected set,
$\zeta_{\L\cup\partial\L} \in \Z^{\{b \in \vec L \mid b \subset (\L \cup \partial\L)\}}$ and $Z^\Lambda_{w,s}$ is a normalization constant.\\
\end{thm}
	The measures $\nu_{w,s}$ will be called pinned gradient measures.

If $q$-periodic boundary law and the underlying potential are translation invariant then it is possible to obtain
probability measure $\nu$ on the gradient space by mixing the pinned gradient measures:

\begin{thm}\cite{KS}	
Let a $q$-periodic boundary law $l$ and  its gradient interaction potential are translation invariant.
Let $\Lambda \subset V$ be any finite connected set and let $w\in \Lambda$ be any vertex. Then the measure $\nu $ with marginals given by
\begin{equation}
\nu (\eta_{\L\cup\partial \L} = \zeta_{\L\cup\partial\L}) = Z_\L \ \left(\sum_{s\in\Z_q}  \prod_{y \in \partial\L} l \big(s + \sum_{b \in \Gamma(w,y)} \zeta_{b}\big)  \right)\prod_{b \cap \L \neq \emptyset} Q(\zeta_b),
\end{equation}
where $Z_\L$ is a normalisation constant, defines a translation invariant gradient Gibbs measure on $\Omega^\nabla$.
\end{thm}

{\bf SOS model.} The (formal) Hamiltonian of the SOS model is
\begin{equation}\label{nu1}
 H(\omega)=-J\sum_{\langle x,y\rangle \in L}
|\omega_x-\omega_y|, \ \ \omega \in\Omega,
\end{equation}
where $J \in \mathbb R_+$ is a constant.

In \cite{HKLR}, using Theorem 3  some gradient Gibbs measures are found.

 Let $\beta>0$ be inverse temperature and $\theta:= \exp(-J\beta)<1$.
 The transfer operator $Q$ then reads $Q(i-j)=\theta^{\vert i-j \vert }$ for any $i,j \in \mathbb{Z}$,
 and a translation invariant boundary law, denoted by $\mathbf z$, is any positive function on $\mathbb{Z}$
 solving the consistency equation, whose values we will denote by $z_i$ instead of $z(i)$.
 By definition of the boundary law it is only unique up to multiplication with any positive prefactor.
 Hence we may choose this constant in a way such that we have $z_0=1$.

 Set $\mathbb{Z}_0:= \mathbb{Z} \setminus \{0\}$. Then the boundary law equation (for translation-invariant case, i.e.
 $l_b\equiv l$, for all $b\in L$) reads
\begin{equation}\label{nu11}
z_i=\left({\theta^{|i|}+
\sum_{j\in \mathbb Z_0}\theta^{|i-j|}z_j
\over
1+\sum_{j\in \mathbb Z_0}\theta^{|j|}z_j}\right)^k, \ \ i\in\mathbb Z_0.
 \end{equation}

Let $\mathbf z(\theta)=(z_i=z_i(\theta), i\in \mathbb Z_0)$ be a solution to (\ref{nu11}).

Denote $u_i=\sqrt[k]{z_i}$ and assume $u_0=1$.

\begin{pro} \cite{HKLR} If $z_0=1$ (i.e. $u_0=1$) then the equation (\ref{nu11}) is equivalent to the following
\begin{equation}\label{V}
u_i^k={u_{i-1}+u_{i+1}-\tau u_i\over u_{-1}+u_{1}-\tau}, \ \ i\in \mathbb Z,
\end{equation}
where $\tau=\theta^{-1}+\theta$.
\end{pro}

In general, solutions of (\ref{V}) are not known. But in class of periodic solutions, some results are obtained.
The following theorem is proved for $k=2$ and $4$-periodic boundary laws:
 	
\begin{thm} \cite{HKLR} For the SOS model (\ref{nu1}) on the binary tree (i.e. $k=2$) with parameter $\tau=\theta+\theta^{-1}$ the following assertions hold
	\begin{itemize}
		\item[1.] If $\tau \leq 4$ then there is precisely one GGM associated to a 4-periodic  boundary law.
		\item[2.] If $4< \tau \leq 6$  then there are precisely two GGMs.
		\item[3.] If $6<\tau < 2+2\sqrt{5}$  then there are precisely three GGMs.
		\item[4.] If $\tau \geq  2+2\sqrt{5}$  then there are precisely four such measures.
	\end{itemize}
\end{thm}

The following theorem is proved for any $k\geq 2$ and $3$-periodic boundary laws.

Denote
$$\tau_0:={2k+1\over k-1}.$$

\begin{thm} \cite{HKLR} For the SOS-model on the $k$-regular tree, $k \geq 2$, with parameter $\tau$ there is $\tau_c$ such that
$0<\tau_c<\tau_0$ and the following holds:
\begin{itemize}
	\item[1.] If $\tau<\tau_c$ then there is no any GGM corresponding to
	a nontrivial 3-periodic boundary.
	\item[2.] At $\tau=\tau_c$ there is a unique GGM corresponding to a
	nontrivial $3$-periodic boundary law.
	\item[3.] For $\tau>\tau_c$, $\tau\ne \tau_0$ (resp. $\tau=\tau_0$) there are exactly
	two such (resp. one) GGMs.
\end{itemize}	

 The GGMs described above are all different from the GGMs mentioned in Theorem 4.
\end{thm}

{\bf General case.}
 Assume that the transfer operator $\{Q_b\}_{b\in \mathbb L}$, defined in (\ref{Qd}), is summable, i.e.
$$\sum_{i\in \mathbb Z}Q_b(i) <\infty \ \ \mbox{for all} \ \  b\in \mathbb L.$$

The following is the main result of \cite{HKa}:

\begin{thm} For any summable $Q$ and any degree $k\geq 2$ there is a finite period
$q_0(k)$ such that for all $q\geq q_0(k)$ there are GGMs of
period $q$ which are not translation invariant.
\end{thm}

Moreover, in \cite{HKb} the authors provided general conditions in terms of the
relevant $p$-norms of the associated transfer operator $Q$ which ensure the existence of
a countable family of proper Gibbs measures. The existence of delocalized GGMs is proved, under natural conditions on $Q$.
This implies coexistence of both types of measures for large classes of models including the SOS-model,
and heavy-tailed models arising for instance for potentials of logarithmic growth.

\section{4-periodic boundary laws for $k\geq 2$}

In this section our goal  is to find solutions of (\ref{V}) which have the form
\begin{equation}\label{up}
u_n=\left\{ \begin{array}{lll}
1, \ \ \mbox{if} \ \ n=2m,\\[2mm]
a, \ \ \mbox{if} \ \ n=4m-1, \ \ m\in \mathbb Z\\[2mm]
b, \ \ \mbox{if} \ \ n=4m+1,
\end{array}
\right.
\end{equation}
where $a$ and $b$ some positive numbers.

Then from (\ref{V}) for $a$ and $b$ we get the following system of equations
\begin{equation}\label{ab}
\begin{array}{ll}
(a+b-\tau)b^k+\tau b-2=0\\[2mm]
(a+b-\tau)a^k+\tau a-2=0.
\end{array}
\end{equation}

The case $k=2$ is fully analyzed in \cite{HKLR} and the following is proved
\begin{pro}\label{pps} For $k=2$ the periodic solutions of the form (\ref{up}) (i.e. solutions of the system (\ref{ab}))
depend on the parameter $\tau=2\cosh(\beta)$ in the following way.
	\begin{enumerate}
		\item If $\tau \leq 4$ then there is a unique solution.
		\item If $4<\tau \leq 6$  then there are exactly two solutions.
		\item If $6<\tau < 2+2\sqrt{5}$  then there are exactly four solutions.
		\item If $\tau \geq 2+2\sqrt{5}$  then there are exactly five solutions.
	\end{enumerate}
	where explicit formula of each solution is found.
\end{pro}

Now we reduce the system (\ref{ab}) to a polynomial equation with one unknown $a$. To do this
from the first (resp. second) equation of (\ref{ab}) find $a$ (resp. $b$):

\begin{equation}\label{ab1}
\begin{array}{ll}
a=f(b):=\tau-b+(2-\tau b)b^{-k}\\[2mm]
b=f(a).
\end{array}
\end{equation}
Thus the system (\ref{ab}) is reduced to
\begin{equation}\label{ab2}
a=f(f(a)).
\end{equation}
Note that solutions of $a=f(a)$ are solutions to (\ref{ab2}) too. It is easy to see that $a=f(a)$ is equivalent to
\begin{equation}\label{ab3}
Q(a):=2a^{k+1}-\tau a^k+\tau a-2=0
\end{equation}

The equation (\ref{ab3}) has the solution $a=1$ independently of the parameters $(\tau, k)$. Dividing both sides of (\ref{ab3}) by $a-1$
we get
\begin{equation}\label{uy22}
2a^k+(2-\tau)(a^{k-1}+a^{k-2}+\dots+a)+2=0.
\end{equation}
The following lemma gives the number of solutions to equation (\ref{uy22}) (compare with Lemma 4.7 in   \cite{HKLR}):
\begin{lemma}\label{l6} For each $k\geq 2$,
	there is exactly one critical value of $\tau$, i.e., $\tau_c=\tau_c(k):=2\cdot {k+1\over k-1}$, such that
	\begin{enumerate}
		\item if $\tau<\tau_c$ then (\ref{uy22}) has no positive solution;
		\item if $\tau=\tau_c$ then the equation has a unique solution $a=1$;
		\item if $\tau>\tau_c$,	then it has exactly two solutions (both different from 1);
			\end{enumerate}
\end{lemma}
\begin{proof}
From (\ref{uy22}) we get
	\[\tau=\psi_k(a):=2+\frac{2(a^k+1)}{a^{k-1}+a^{k-2}+\dots+a}. \]
	We have $\psi_k(a)>2$, $a>0$ and $\psi_k'(a)=0$ is equivalent to
	\begin{equation}\label{ho}
	\sum_{j=1}^{k-1}(k-j)a^{k+j-1}-\sum_{j=1}^{k-1}ja^{j-1}=0.
	\end{equation}
	The last polynomial equation has exactly one positive solution,
	because signs of its coefficients changed only one time,
	and at $a=0$ it is negative, i.e. -1 and at $a=+\infty$ it is positive.
	Moreover, this unique solution is $a=1$, because putting $a=1$ in (\ref{ho}) we get
 $$\sum_{j=1}^{k-1}(k-j)-\sum_{j=1}^{k-1}j=\sum_{j=1}^{k-1}k-2\sum_{j=1}^{k-1}j = k(k-1)-2\cdot {k(k-1)\over 2}=0.$$

 Thus $\psi_k(a)$ has unique minimum
	at $a=1$, and $\lim_{a\to 0}\psi_k(a)=\lim_{a\to +\infty}\psi_k(a)=+\infty$.
Consequently,
	\[\tau_c=\tau_c(k)=\min_{a>0}\psi_k(a)=\psi_k(1)=2\cdot {k+1\over k-1}.\]
	These properties of $\psi_k(a)$ completes the proof.
\end{proof}
\begin{rk} The equation (\ref{uy22}) was also considered in \cite{BH}.
Lemma \ref{l6} improves their result (see Theorem 5.2 of \cite{BH}), because we found explicit
formula for the critical value $\tau_c$.

\end{rk}
Now we want to find solutions of (\ref{ab2}) which are different from solutions of $a=f(a)$ (i.e., $Q(a)=0$).

By simple calculations the equation (\ref{ab2}) can be rewritten as
\begin{equation}\label{ab4}
P(a):=(2-\tau a)[2-\tau a+\tau a^k-a^{k+1}]^k-a^{k^2}[\tau a^{k+1}+(2-\tau^2)a^k+\tau^2a-2\tau]=0.
\end{equation}
Recall that $Q(a)$ divides $P(a)$. Now we shall find  ${P(a)\over Q(a)} $.
It is easy to see that $P(a)$ can be written as
$$P(a)=(2-\tau a)[a^{k+1}-Q(a)]^k-a^{k^2}[2a^k-\tau a^{k+1}+\tau Q(a)]$$
$$=(2-\tau a)\sum_{j=0}^{k}(-1)^{k-j}{k\choose j}a^{(k+1)j}Q^{k-j}(a)-a^{k^2+k}(2-\tau a)-\tau a^{k^2}Q(a)$$
$$=(2-\tau a)\sum_{j=0}^{k-1}(-1)^{k-j}{k\choose j}a^{(k+1)j}Q^{k-j}(a)+(2-\tau a)a^{k^2+k}-a^{k^2+k}(2-\tau a)-\tau a^{k^2}Q(a)$$
$$=(2-\tau a)\sum_{j=0}^{k-1}(-1)^{k-j}{k\choose j}a^{(k+1)j}Q^{k-j}(a)-\tau a^{k^2}Q(a)$$
$$=Q(a)\left((2-\tau a)\sum_{j=0}^{k-1}(-1)^{k-j}{k\choose j}a^{(k+1)j}Q^{k-j-1}(a)-\tau a^{k^2}\right).$$
%
%

Consequently, the equation (\ref{ab2}) in case $Q(a)\ne 0$  is reduced to
\begin{equation}\label{kt}
U(a):=\tau a^{k^2}-(2-\tau a)\sum_{j=0}^{k-1}(-1)^{k-j}{k\choose j}a^{(k+1)j}Q^{k-j-1}(a)=0.
\end{equation}
But $U(a)$ may have some roots coinciding with roots of $Q(a)$. This is result of the following lemma.

\begin{lemma} Let $a=\hat a$ be a root of $Q(a)$. Then $\hat a$ is a root of $U(a)$ iff
$\hat a={2k\over \tau (k-1)}$ and $\tau$ satisfies
\begin{equation}\label{ha}
(k-1)^k\tau^{k+1}-(k-1)2^{k-1}k^k\tau^2+(2k)^{k+1}=0.
\end{equation}
\end{lemma}
\begin{proof} For the root $\hat a$ we have $Q(\hat a)=0$. Therefore from (\ref{kt}), i.e. $U(\hat a)=0$, we get
$$\tau \hat a^{k^2}+(2-\tau \hat a)k \hat a^{k^2-1}=0 \ \ \Leftrightarrow \ \ \tau \hat a+(2-\tau \hat a)k=0 \ \ \Leftrightarrow \ \ \hat a={2k\over \tau (k-1)}.$$

Then $Q(\hat a)=Q\left({2k\over \tau (k-1)}\right)=0$ gives (\ref{ha}).

\end{proof}

\begin{lemma}\label{s3} For each fixed $k\geq 2$ the equation (\ref{ha}) has exactly two
solutions $\tau_1(k)={2k\over k-1}$ and $\tau_2(k)>{2(k+1)\over k-1}$.
\end{lemma}
\begin{proof} It is easy to see that $\tau={2k\over k-1}$ is a solution to (\ref{ha}),
for any $k\geq 2$. Moreover, the corresponding
$\hat a$ is $1$.

Denote $L=(k-1)\tau-2k$
 then from (\ref{ha}) we get
$$0=(L+2k)^{k+1}-2^{k-1}k^k(L+2k)^2+(k-1)(2k)^{k+1}$$
$$=\sum_{j=0}^{k+1}{k+1\choose j}L^{k+1-j}(2k)^j-2^{k-1}k^kL^2-(2k)^{k+1}L-2^{k+1}k^{k+2}+(k-1)(2k)^{k+1}
$$ $$=\sum_{j=0}^{k}{k+1\choose j}L^{k+1-j}(2k)^j+(2k)^{k+1}-2^{k-1}k^kL^2-(2k)^{k+1}L-2^{k+1}k^{k+2}+(k-1)(2k)^{k+1}$$
$$=\sum_{j=0}^{k}{k+1\choose j}L^{k+1-j}(2k)^j-2^{k-1}k^kL^2-(2k)^{k+1}L$$
$$=L\left(\sum_{j=0}^{k}{k+1\choose j}L^{k-j}(2k)^j-2^{k-1}k^kL-(2k)^{k+1}\right)$$
$$=L\left(\sum_{j=0}^{k-2}{k+1\choose j}L^{k-j}(2k)^j+{k(k-1)\over 2}L-(k-1)(2k)^{k}\right).$$
Thus $L=0$ (i.e. $\tau=\tau_1(k)$) or
\begin{equation}\label{L}\sum_{j=0}^{k-2}{k+1\choose j}L^{k-j}(2k)^j+{k(k-1)\over 2}L-(k-1)(2k)^{k}=0.
\end{equation}
By Lemma \ref{l6} we know that $Q(a)=0$ has a solution different
from 1 iff $\tau>\tau_c={2(k+1)\over k-1}$. Therefore $L=(k-1)\tau-2k>(k-1)\cdot {2(k+1)\over k-1}-2k=2>0$.
The polynomial equation (\ref{L}) has exactly one positive solution, denoted by $L^*$,
	because signs of its coefficients changed only one time. Moreover, we have $L^*>2$, because at $L=2$ the LHS of (\ref{L}) is negative:
$$2^k[(k+1)^{k+1}-(3k+1)k^k]<0$$ and at $L=+\infty$ it is positive. Thus $\tau_2(k)={L^*+2k\over k-1}$.
\end{proof}
\begin{ex} For $k=2,3,4$ we have
$$\tau_1(2)=4, \ \ \tau_2(2)=2+2\sqrt{5}.$$
$$\tau_1(3)=3, \ \ \tau_2(3)=3\sqrt{2}.$$
$$\tau_1(4)=8/3, \ \ \tau_2(4)\approx 3.497.$$

\end{ex}

\begin{rk} It seems impossible to solve equation (\ref{kt}) for each $k\geq 2$. But one can use numerical
methods to give some its solutions for concrete values of parameters.
Therefore, one can try to solve it for small values of $k$. In \cite{HKLR} the case $k=2$ is fully analyzed.
Below we shall consider the case $k=3$. Cases $k\geq 4$ remains open.
\end{rk}

\begin{figure}[h!]
\vspace{-.5pc} \centering
\includegraphics[width=7.2cm]{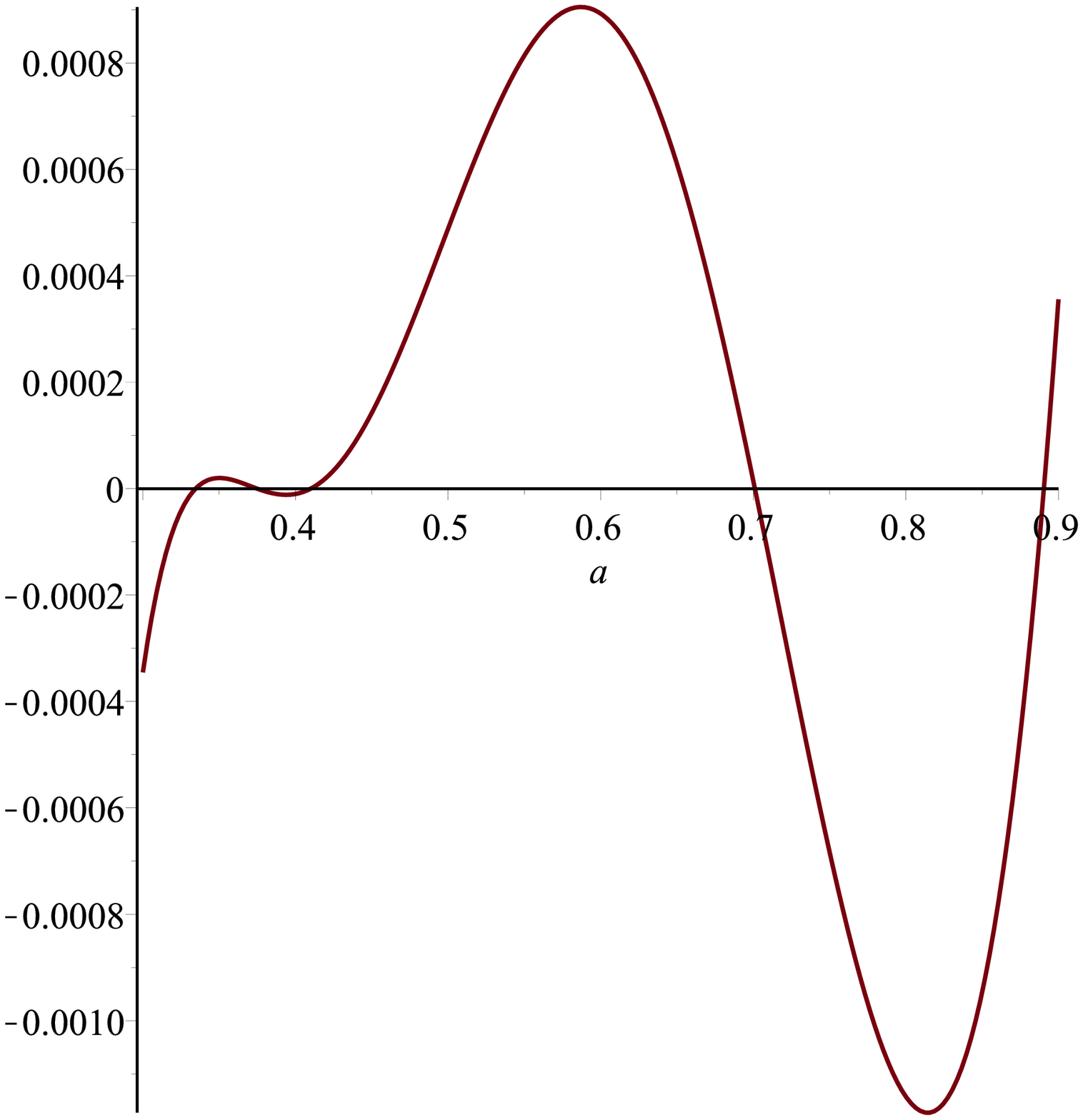}
\includegraphics[width=7.2cm]{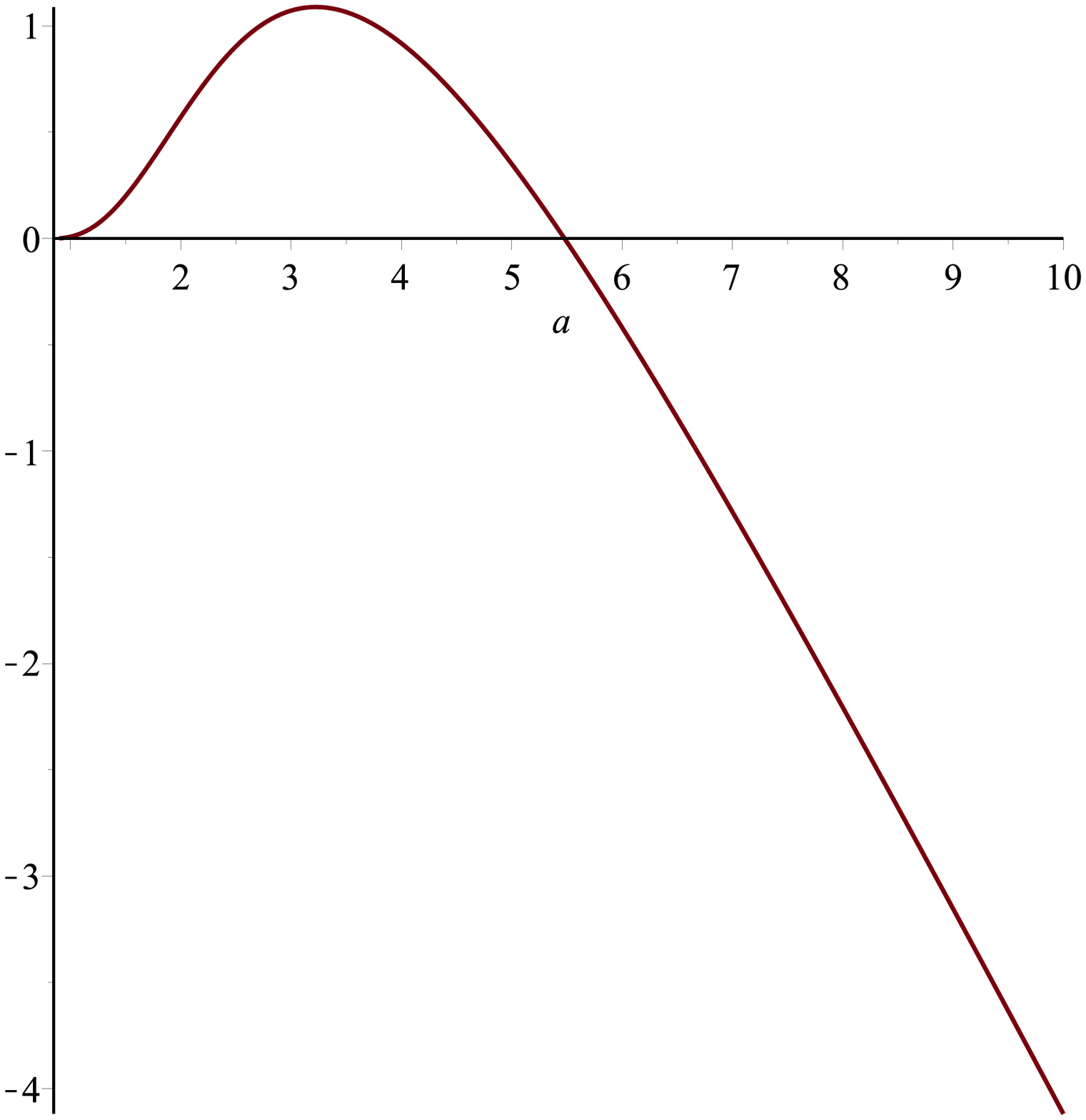}
\caption{Graph of function $Y(x,6)-x$, for $x\in (0.3,0.9)$ and $x\in (0.9,10)$. Hence, $x=Y(x,6)$ has 6 (the maximal number) positive solutions.}\label{k3}
\end{figure}

{\bf Case $k=3$}. In this case the equation (\ref{kt}) has the form

$$g(a,\tau):=(\tau^2+2)a^8-(\tau^2+2)\tau a^7+\tau^2 a^6+2(\tau^2+2)\tau a^5$$
\begin{equation}\label{g}-4(2\tau^2+1)a^4-(\tau^2-8)\tau a^3+6\tau^2a^2-12\tau a+8=0.
\end{equation}
It is well known (see \cite{Pra}, p.28) that the number of positive
roots of the polynomial (\ref{g}) does not exceed the number of sign
changes of its coefficients.
Since $\tau>2$, the number of positive roots of
the polynomial (\ref{g}) is at most $6$. Numerical analysis shows that
for some values of $\tau$ there are exactly 6 solutions (see Fig.\ref{k3}).
Indeed, rewrite (\ref{g}) as
$$a=Y(a,\tau):={(\tau^2+2)\tau a^7+4(2\tau^2+1)a^4+(\tau^2-8)\tau a^3
+13\tau a-8\over(\tau^2+2)a^7+\tau^2 a^5+2(\tau^2+2)\tau a^4+6\tau^2 a+\tau}.$$
Note that, for fixed $\tau>2$, the function $Y(a, \tau)$ is continuous with respect
to both arguments $a>0$, $\tau>2$ and is a bounded function.
Moreover, $Y(0,\tau)=-{8\over \tau}$ and $Y(+\infty, \tau)=\tau$.

\begin{rk} \label{rk} For $k=3$ one can explicitly find all positive solutions of (\ref{ab3}), i.e.
$$a_1=1, \ \ 0<a_2={1\over 4}(\tau-\sqrt{\tau^2-16})<1, \ \ 0<a_3={1\over 4}(\tau+\sqrt{\tau^2-16})>1.$$
Moreover, $a_2a_3=1$.
\end{rk}

%
In case $a\ne b$, we do not know explicit solutions of (\ref{g}).
Since $b=f(a)$ may be negative for some positive solutions $a$,
we have to check positivity of $b$ for each positive $a$. To avoid this difficulty, in case $k=3$ and $a\ne b$, we solve (\ref{ab}) as
follows.

We rewrite the system of equations (\ref{ab}) for the case $k=3$.
\begin{equation}\label{eq1.1}\left\{
  \begin{array}{ll}
    (a+b-\tau)a^3+\tau a-2=0; \\
    (a+b-\tau)b^3+\tau b-2=0.
  \end{array}
\right.
\end{equation}
\begin{lemma}\label{new} For the system (\ref{eq1.1})  there are critical
values $\tau^{(1)}_{cr}\approx 3.13039$ and $\tau^{(2)}_{cr}\approx 4,01009$ of $\tau$ such that the following assertions hold
\begin{enumerate}
\item If $\tau\in [2, \tau^{(1)}_{cr}]$ then there is no any solution.
\item If $\tau\in (\tau^{(1)}_{cr}, \tau^{(2)}_{cr}]$ then there is precisely one solution.
\item If $\tau\in(\tau^{(2)}_{cr},\infty)$ then there are precisely two solution to (\ref{eq1.1}).
\end{enumerate}
\end{lemma}
\begin{proof}
We add the first and second equations  of (\ref{eq1.1}), i.e.,
$$(a+b-\tau)(a^3+b^3)+\tau(a+b)-4=0.$$
If $a+b-\tau=0$ then there is one solution to (\ref{eq1.1}), i.e., $a=b=\frac{2}{\tau}$. We have to find $(a,b), a\neq b$ solutions, so we can suppose $a+b-\tau\neq 0$. Consequently,
$$a^3+b^3=\frac{4-\tau(a+b)}{a+b-\tau}.$$
From the last equality, one gets
\begin{equation}\label{eq1.2} 3ab=(a+b)^2+\frac{\tau(a+b)-4}{(a+b)(a+b-\tau)}.
\end{equation}
Now, we subtract the second equation of (\ref{eq1.1}) from the first one. Then
$$(a+b-\tau)(a^3-b^3)+\tau(a-b)=0.$$
Since $a\neq b$, both sides can be divided by $a-b$ and we obtain the following
\begin{equation}\label{eq1.3} ab=(a+b)^2+\frac{\tau}{a+b-\tau}.
\end{equation}
Let $a+b=x$. Then by (\ref{eq1.2}) and (\ref{eq1.3}), we have
$$2x^2+\frac{3\tau}{x-\tau}=\frac{\tau x-4}{x(x-\tau)}.$$
The last equation can be written as
\begin{equation}\label{eq1.4} x^4-\tau x^3+\tau x+2=0.\end{equation}
By Ferrari's method for solving a quartic equation, the equation (\ref{eq1.4}) can be written as
\begin{equation}\label{eq1.5} \left(x^2-\frac{\tau}{2}x+\frac{c(\tau)}{2}\right)^2-
\left[\left(\frac{\tau^2}{4}+c(\tau)\right)x^2-\left(\frac{\tau c(\tau)}{2}+\tau\right)x+\left(\frac{c^2(\tau)}{4}-2\right)\right]=0,\end{equation}
where $c(\tau)$ is a real root of the following polynomial
$$P(z):=z^3-(\tau^2+8)z-3\tau^2=0.$$
Denote
$$F(\tau) = \dfrac {-(\tau^2+8)} {3}, \quad R(\tau)= \dfrac {3\tau^2} {2}.$$
To find a certain view of $c(\tau)$, we shall find a real root of $P(z)$. By Cardano's formula, roots of $P(z)$ are:
$$z_1(\tau) =S(\tau)+T(\tau),$$
$$z_2(\tau)= - \dfrac {S(\tau) + T(\tau)} 2 + \dfrac {i \sqrt 3} 2 (S(\tau)- T(\tau)),$$
$$z_3(\tau)= - \dfrac {S(\tau) + T(\tau)} 2 - \dfrac {i \sqrt 3} 2 (S(\tau)- T(\tau)),$$
where
$$S(\tau)= \sqrt [3] {R(\tau)+ \sqrt {F^3(\tau)+R^2(\tau)} }, \quad T(\tau)= \sqrt [3] {R(\tau)- \sqrt {F^3(\tau)+ R^2(\tau)} }.$$
It's known that the expression $D(\tau)=F^3(\tau)+R^2(\tau)$ is called the discriminant of the equation.

 \begin{itemize}
\item If $D(\tau)>0$, then one root is real and two are complex conjugates.
\item If $D(\tau)=0$, then all roots are real, and at least two are equal.
\item If $D(\tau)<0$, then all roots are real and unequal.
\end{itemize}
We have
$$D(\tau)=F^3(\tau)+R^2(\tau)=-\frac{1}{27}\tau^6+\frac{49}{36}\tau^4-\frac{64}{9}\tau^2-\frac{512}{27}.$$
Note that there are two positive solutions of $D(\tau)$ on $[2,+\infty)$ i.e., $\tau_1\approx 2.994$ and $\tau_2\approx 5.45$ and $D(\tau)<0$ for $\tau\in[2, \tau_1)\cup (\tau_2, +\infty)$ and $D(\tau)>0$ for $\tau\in(\tau_1, \tau_2).$
For all cases, $S(\tau)+T(\tau)$ is a real solution and we can choose $c(\tau)$ as $S(\tau)+T(\tau)$. In addition, $c(\tau)=S(\tau)+T(\tau)>0$.
The equation (\ref{eq1.5}) can be written as
$$\left(x^2+\left(\sqrt{\frac{\tau^2}{4}+c(\tau)}-\frac{\tau}{2}\right)x+\frac{c(\tau)}{2}-\sqrt{\frac{c^2(\tau)}{4}-2}\right)\times$$
$$\times\left(x^2-\left(\sqrt{\frac{\tau^2}{4}+c(\tau)}+\frac{\tau}{2}\right)x+\frac{c(\tau)}{2}+\sqrt{\frac{c^2(\tau)}{4}-2}\right)=0.$$
It's easy to check that
$$x^2+\left(\sqrt{\frac{\tau^2}{4}+c(\tau)}-\frac{\tau}{2}\right)x+\frac{c(\tau)}{2}-\sqrt{\frac{c^2(\tau)}{4}-2}>0, \ x\in \mathbb{R}.$$
Hence, from the last equality we obtain
$$x_{1,2}(\tau)=\frac{1}{2}\left[\sqrt{\frac{\tau^2}{2}+c(\tau)}+\frac{\tau}{2}\pm \sqrt{\left(\sqrt{\frac{\tau^2}{2}+c(\tau)}+\frac{\tau}{2}\right)^2-2\left(c(\tau)+\sqrt{c^2(\tau)-8}\right)}\right].$$
$$\left(\sqrt{\frac{\tau^2}{2}+c(\tau)}+\frac{\tau}{2}\right)^2-2\left(c(\tau)+\sqrt{c^2(\tau)-8}\right)\geq 0\ \Leftrightarrow \ \tau\geq\tau_1\approx 2.994.$$
From the equation (\ref{eq1.3})
$$ab=x^2_{i}(\tau)+\frac{\tau}{x^2_{i}(\tau)-\tau}, \ i\in \{1,2\}.$$
Namely,
$$a(x_{i}(\tau)-a)=x^2_{i}(\tau)+\frac{\tau}{x^2_{i}(\tau)-\tau}.$$
Consequently,
$$a_{1,2}^{(i)}(\tau)=\frac{x_{i}(\tau)\pm\sqrt{x^2_{i}(\tau)-4\left(x^2_i(\tau)+\frac{\tau}{x_{i}(\tau)-\tau}\right)}}{2}.$$
From numerical analysis, $$x^2_{1}(\tau)-4\left(x^2_1(\tau)+\frac{\tau}{x_{1}(\tau)-\tau}\right)< 0, \ \tau\in[\tau_1,+\infty).$$
Also,
$$x^2_{2}(\tau)-4\left(x^2_2(\tau)+\frac{\tau}{x_{2}(\tau)-\tau}\right)\geq 0, \ \tau\in[\tau^{(1)}_{cr},+\infty), \ \tau^{(1)}_{cr}\approx 3.13039.$$
Hence, we have only two cases $a\in \{a_1^{(2)}(\tau), a_2^{(2)}(\tau)\}$. When $a_1^{(2)}(\tau)$ (resp $a_2^{(2)}(\tau)$) belongs to the interval $(0, x_{2}(\tau))$ then $b_1^{(2)}(\tau)$ (resp. $b_2^{(2)}(\tau)$) is also positive. Again we use numerical analysis and obtain the following results: if $\tau\in[2,\tau^{(1)}_{cr}]$ then $a_2^{(2)}(\tau)\geq x_2(\tau)$, i.e., the equation $(\ref{eq1.1})$ has no any positive solution such that $a\neq b$. Also, for all $\tau\in(\tau^{(1)}_{cr}, \tau^{(2)}_{cr}]$ we have $a_2^{(2)}(\tau)<x_2(\tau)$,  i.e., the equation $(\ref{eq1.1})$ has exactly one positive solution with $a\neq b$. Let $\tau\in(\tau^{(2)}_{cr}, +\infty)$, then $a_1^{(2)}(\tau)<x_2(\tau)$ and in this case the equation $(\ref{eq1.1})$ has exactly two positive solutions with $a\neq b$, where $\tau^{(2)}_{cr}\approx 4,01009$.
\end{proof}
Denote by $\mathcal N_k(\tau)$ the number of positive roots of (\ref{eq1.1}).
Then (for $k=3$) by Lemma \ref{l6} (where $\tau_c(3)=4$), Lemma \ref{s3}, Lemma \ref{new}
we obtain the following formula
 \begin{equation}\label{nk}\mathcal N_3(\tau)=\left\{\begin{array}{llllllllll}
 1, \ \ \mbox{if} \ \ \tau\in (2, \tau_{cr}^{(1)}]\\[2mm]
 2, \ \ \mbox{if} \ \ \tau\in (\tau_{cr}^{(1)},4]\\[2mm]
 3, \ \ \mbox{if} \ \ \tau\in (4, \tau_{cr}^{(2)}]\cup\{3\sqrt{2}\}\\[2mm]
 4, \ \ \mbox{if} \ \ \tau\in (\tau_{cr}^{(2)}, +\infty)\setminus\{3\sqrt{2}\},
 \end{array}\right.
 \end{equation}
where $\tau^{(1)}_{cr}\approx 3.13039$,  $\tau^{(2)}_{cr}\approx 4,01009$.

In \cite{HKLR} some statements on identifiability of GGM with respect to the class of boundary laws are proven.
In particular, for 4-periodic case the following is known.
\begin{lemma}\label{4p}\cite{HKLR} 	Consider any 4-periodic boundary law constructed by $a,b$ given in (\ref{up})
	and denote the associated GGM by $\nu^{(a,b)}$. Let $(a_1,b_1) ,(a_2,b_2)$ be two such boundary laws. If $\nu^{(a_1,b_1)}=\nu^{(a_2,b_2)}$ then necessarily
	\begin{equation*}
	\begin{split}
	&a_1+b_1 = a_2+b_2 \quad \text{or } \cr
	&(a_1+b_1)(a_2+b_2) = 4.
	\end{split}
	\end{equation*}
\end{lemma}

Based on formula (\ref{nk}), Remark \ref{rk} and Lemma \ref{4p} we conclude the following
\begin{thm} For the SOS model (\ref{nu1}) on the Cayley tree of order $k=3$
there are critical values $\tau^{(1)}_{cr}\approx 3.13039$,  $\tau^{(2)}_{cr}\approx 4,01009$ such that the following assertions hold
	\begin{enumerate}
		\item If $\tau \leq \tau_{cr}^{(1)}$ then there is precisely one GGM associated to a boundary law of the type \eqref{up}.
		\item If $\tau\in (\tau_{cr}^{(1)},4]$ then there are precisely two such GGMs.
		\item If $\tau\in (4, \tau_{cr}^{(2)}]\cup\{3\sqrt{2}\}$  then there are at most three such GGMs.
		\item If $\tau\in (\tau_{cr}^{(2)}, +\infty)\setminus\{3\sqrt{2}\}$  then there are at
most four such measures associated to boundary laws of the type \eqref{up}.
	\end{enumerate}
\end{thm}

\section{SOS model with an external field}

\subsection{The boundary law equation in case of non-zero external field.}

%
%
In this section for  $\sigma_n:x\in V_n\mapsto \sigma(x)\in \mathbb Z$, consider Hamiltonian
of SOS model with external field  $\Phi:\mathbb Z\to \mathbb R$, i.e.,
$$H(\sigma_n)=-J\sum_{\langle x,y\rangle: \atop x,y\in V_n}
|\sigma(x)-\sigma(y)|+\sum_{x\in V_n}\Phi(\sigma(x)).$$
Denote $h(i)=\exp(\Phi(i))$, $i\in \mathbb Z$.

Then the equation for translation-invariant boundary laws has the following form
\begin{equation}\label{di1}
z_i=\frac{h(i)}{h(0)}\left({\theta^{|i|}+
\sum_{j\in \mathbb Z_0}\theta^{|i-j|}z_j
\over
1+\sum_{j\in \mathbb Z_0}\theta^{|j|}z_j}\right)^k, \ \ i\in\mathbb Z_0.
 \end{equation}
Note that this equation coincides with (\ref{nu11}) for $ \frac{h(i)}{h(0)}\equiv 1$.

Let $\mathbf z(\theta)=(z_i=z_i(\theta)>0, i\in \mathbb Z_0)$ be a solution to (\ref{di1}).   Denote
\begin{equation}\label{lr}
l_i\equiv l_i(\theta)=\sum_{j=-\infty}^{-1}\theta^{|i-j|}z_j, \ \
r_i\equiv r_i(\theta)=\sum_{j=1}^{\infty}\theta^{|i-j|}z_j, \ \ i\in\mathbb Z_0.
\end{equation}
It is clear that each $l_i$ and $r_i$ can be a finite positive number or $+\infty$.

\begin{lemma}\label{l1} \cite{HKLR} For each $i\in \mathbb Z_0$ we have
\begin{itemize}
\item $l_i<+\infty$ if and only if $l_0<+\infty$;

\item $r_i<+\infty$ if and only if $r_0<+\infty$.
\end{itemize}
\end{lemma}


\begin{pro} \label{pps: 1} Assume $h(0)=1$.
	A vector $\mathbf z=(z_i,i\in \mathbb Z)$, with $z_0=1$,  is a solution to (\ref{di1})
if and only if for $s_i=\sqrt[k]{{z_i\over h(i)}}$ the following holds
	\begin{equation}\label{Va}
	h(i)s_i^k=\frac{{s_{i-1}+s_{i+1}-\tau s_i}}{s_{-1}+s_{1}-\tau}, \ \ i\in \mathbb Z,
	\end{equation}
	where $\tau=\theta^{-1}+\theta=2\cosh(\beta)$.
\end{pro}
\begin{proof} (cf. with the proof of Proposition 4.3 of \cite{HKLR}).
Take some $C>0$ and denote
$$v_i=C\cdot h^{1/k}(i)\left(\theta^{|i|}+
\sum_{j\in \mathbb Z_0}\theta^{|i-j|}z_j\right), \ \ i\in \mathbb Z.$$

Then from (\ref{di1}) we get $z_i=\left({v_i\over v_0}\right)^k$ and consequently,
\begin{equation}\label{di2}
\left({v_i\over v_0}\right)^k=\frac{h(i)}{h(0)}\left({\theta^{|i|}+
\sum_{j\in \mathbb Z_0}\theta^{|i-j|}\left({v_j\over v_0}\right)^k
\over
1+\sum_{j\in \mathbb Z_0}\theta^{|j|}\left({v_j\over v_0}\right)^k}\right)^k, \ \ i\in\mathbb Z_0.
 \end{equation}
From (\ref{di2}) we obtain
$$v_i=C\cdot \sqrt[k]{h(i)}\left(\sum_{j=1}^{+\infty}\theta^jv_{i-j}^k+v_i^k+ \sum_{j=1}^{+\infty}\theta^jv_{i+j}^k\right), \ \ i \in \mathbb{Z}.
$$
By the last equality we get
$${v_{i-1}\over \sqrt[k]{h(i-1)}}+{v_{i+1}\over \sqrt[k]{h(i+1)}}-\tau\cdot {v_{i}\over \sqrt[k]{h(i)}}=C\cdot (\theta-{1\over \theta}) v^k_i.$$
This equality by the notation $s_i=\sqrt[k]{{z_i\over h(i)}}$ gives (\ref{Va}). Conversely, from (\ref{Va}) one gets (\ref{di1}).
\end{proof}
\subsection{A case of 4-periodic external field for $k=2$.}
Here we shall find solutions of (\ref{Va}) for external field
\begin{equation}\label{uh}
h(i)=\left\{ \begin{array}{lll}
1, \ \ \mbox{if} \ \ i=2m,\\[2mm]
h_1, \ \ \mbox{if} \ \ i=4m-1, \ \ m\in \mathbb Z\\[2mm]
h_2, \ \ \mbox{if} \ \ i=4m+1,
\end{array}
\right.
\end{equation}
where $h_1$ and $h_2$ are some positive numbers and
a solution which has the form

\begin{equation}\label{upa}
u_n=\left\{ \begin{array}{lll}
1, \ \ \mbox{if} \ \ n=2m,\\[2mm]
a, \ \ \mbox{if} \ \ n=4m-1, \ \ m\in \mathbb Z\\[2mm]
b, \ \ \mbox{if} \ \ n=4m+1,
\end{array}
\right.
\end{equation}
where $a$ and $b$ some positive numbers.

Then from (\ref{Va}) for $a$ and $b$ we get the following system of equations
\begin{equation}\label{abd}
\begin{array}{ll}
(a+b-\tau)h_2b^k+\tau b-2=0\\[2mm]
(a+b-\tau)h_1a^k+\tau a-2=0.
\end{array}
\end{equation}

For simplicity we consider the case $k=2$ and $h_1=h_2=h$.
In this case, subtracting from the first equation of the system (\ref{abd}) the second one we get
\[(b-a)[h(a+b)^2-h\tau(a+b)+\tau]=0.\]
This gives three possible cases:
\begin{equation}
\label{b}
a=b,  \ \ \mbox{and} \ \  a+b=\frac{1}{2h}(h\tau\pm \sqrt{h\tau(h\tau-4)}) \ \ \mbox{for} \ \ h\tau\geq 4.
\end{equation}

 \textbf{Case $a=b$}. In this case from the first equation of (\ref{abd}) we get
	\begin{equation}
	\label{a3}
	2ha^3-h\tau a^2+\tau a-2=0.
	\end{equation}
\begin{lemma}\label{lp} Any real solution of (\ref{a3}) is positive.
\end{lemma}
\begin{proof} Since $h>0$, $\tau>2$, if $a\leq 0$ then LHS of (\ref{a3}) is strictly negative.
\end{proof}
Rewrite the cubic equation (\ref{a3}) as
$$
a^{3}+\a a^{2}+\b a+\c=0
$$
where $\a=-\tau/2$, $\b=\tau/2h$, $\c=-1/h$.

Let
$$
p=\b-\frac{\a^{2}}{3} \quad \text { and } \quad q=\frac{2 \a^{3}}{27}-\frac{\a \b}{3}+\c
$$
Then the discriminant $\Delta$ of the cubic equation is
$$
\Delta=\frac{q^{2}}{4}+\frac{p^{3}}{27}.
$$

\begin{lemma} The following assertions hold
\begin{itemize}
\item[Case: $\Delta>0 .$] In this case there is only one \textbf{positive} real solution. It is
$$
a_1=\left(-\frac{q}{2}+\sqrt{\Delta}\right)^{\frac{1}{3}}+\left(-\frac{q}{2}-\sqrt{\Delta}\right)^{\frac{1}{3}}+\frac{\tau}{6}
$$

\item[Case: $\Delta=0 .$] In this case there are two \textbf{positive} real solutions. These roots are
$$
a_{1}=-2\left(\frac{q}{2}\right)^{\frac{1}{3}}+\frac{\tau}{6} \quad \text { and } \quad a_{2}=a_{3}=\left(\frac{q}{2}\right)^{\frac{1}{3}}+\frac{\tau}{6}
$$

\item[Case: $\Delta<0 .$] In this case $-p>0$ and there are three \textbf{positive} real solutions:
$$
\begin{array}{l}
a_{1}=\frac{2}{\sqrt{3}} \sqrt{-p} \sin \left(\frac{1}{3} \sin ^{-1}\left(\frac{3 \sqrt{3} q}{2(\sqrt{-p})^{3}}\right)\right)+\frac{\tau}{6} \\
a_{2}=-\frac{2}{\sqrt{3}} \sqrt{-p} \sin \left(\frac{1}{3} \sin ^{-1}\left(\frac{3 \sqrt{3} q}{2(\sqrt{-p})^{3}}\right)+\frac{\pi}{3}\right)+\frac{\tau}{6} \\
a_{3}=\frac{2}{\sqrt{3}} \sqrt{-p} \cos \left(\frac{1}{3} \sin ^{-1}\left(\frac{3 \sqrt{3} q}{2(\sqrt{-p})^{3}}\right)+\frac{\pi}{6}\right)+\frac{\tau}{6}
\end{array}
$$
\end{itemize}
\end{lemma}
\begin{proof} The conditions of existence of real solutions are well-known\footnote{https://en.wikipedia.org/wiki/Cubic$_-$equation}. The positivity of each solution follows from Lemma \ref{lp}.
\end{proof}


\textbf{Case  $a+b=\frac{1}{2h}(h\tau+ \sqrt{h\tau(h\tau-4)})$}.
	In this case from the second equation of (\ref{abd}) we get
	\[(h\tau-\sqrt{h\tau(h\tau-4)})a^2-2\tau a +4=0.\]
	Note that this equation has only positive real solutions.
Moreover, one can see that if
$$(\tau, h)\in A:=\left\{(\tau, h)\in \mathbb R^2_+ \,: \,h\geq \left[\begin{array}{ll}
{\tau^3\over 8(\tau^2-8)}, \ \ \mbox{if} \ \ 2\sqrt{2}<\tau<4\\[2mm]
{4\over \tau}, \ \ \mbox{if} \ \ \tau\geq 4
\end{array}\right.\right\}$$
then the quadratic equation has the following positive solutions
	\[a_4=\frac{\tau-\sqrt{\tau^2-4h\tau+4\sqrt{h\tau(h\tau-4)}}}{h\tau-\sqrt{h\tau(h\tau-4)}}, \ \ a_5=\frac{\tau+\sqrt{\tau^2-4h\tau+4\sqrt{h\tau(h\tau-4)}}}{h\tau-\sqrt{h\tau(h\tau-4)}}. \]

	Using (\ref{b}) we get $b_4=a_5$ and $b_5=a_4$.

\textbf{Case  $a+b=\frac{1}{2h}(h\tau- \sqrt{h\tau(h\tau-4)})$}.
	In this case we obtain
	\begin{equation*}
	(h\tau+ \sqrt{h\tau(h\tau-4)})a^2-2\tau a+4=0
	\end{equation*}
	which
	for
$$(\tau, h)\in B:=\left\{(\tau, h)\in \mathbb R^2_+ \,: \, \tau\geq 4, \, {4\over \tau}\leq h\leq {\tau^3\over 8(\tau^2-8)}\right\}$$ has the following solutions
	\[a_6=\frac{\tau-\sqrt{\tau^2-4h\tau-4\sqrt{h\tau(h\tau-4)}}}{h\tau+\sqrt{h\tau(h\tau-4)}}, \ \ a_7=\frac{\tau+\sqrt{\tau^2-4h\tau-4\sqrt{h\tau(h\tau-4)}}}{h\tau+\sqrt{h\tau(h\tau-4)}}. \]
	Using (\ref{b}) we get $b_6=a_7$ and $b_7=a_6$. Clearly all of these solutions are positive.

Denote by $\mu_i$ the gradient Gibbs measure corresponding to solution $(a_i, b_i)$, $i=1,2,\dots,7$.

Thus depending on the values $(\tau, h)$ related to $\Delta$ and sets $A$, $B$ we have the following result.

\begin{thm} For the SOS model with 4-periodic external field there are up to seven 4-periodic gradient Gibbs measures $\mu_i$, $i=1,\dots,7$.

\end{thm}

\section*{ Acknowledgements}

 The author thanks C. K\"ulske for helpful discussions.

\end{document}